\documentclass[a4paper,11pt]{article}

\usepackage{jheppub} 

\usepackage[T1]{fontenc} 
 \usepackage{bm}
\def\Re{\relax{\textbf{Re}{}}}                            
\newcommand{\Ds}{\displaystyle}                           
 \renewcommand{\thesection}{\arabic{section}}

\renewcommand{\thetable}{\arabic{table}}
\renewcommand{\thefigure}{\arabic{figure}}

\renewcommand{\tablename}{Table}
\renewcommand{\figurename}{Figure}

\newcommand{\be}{\begin{equation}}\newcommand{\ee}{\end{equation}}%
\newcommand{\bd}{\begin{displaymath}}\newcommand{\ed}{\end{displaymath}}
\newcommand{\bit}{\begin{itemize}}                        
 \newcommand{\eit}{\end{itemize}}                         
\newcommand{\ben}{\begin{enumerate}}                      
 \newcommand{\een}{\end{enumerate}}                       
\newcommand{\baa}{\begin{array}{lll}}                     
 \newcommand{\eaa}{\end{array}}                           
\newcommand{\ba}{\begin{eqnarray}}                        
 \newcommand{\ea}{\end{eqnarray}}                         
\newcommand{\gev}[1]{\relax\ifmmode{\text{GeV}^{#1}}      
                     \else{GeV$^{#1}${ }}\fi}             
\newcommand{\Gev}{\relax\ifmmode{\text{GeV}}              
                     \else{GeV{ }}\fi}                    
\newcommand{\Mev}{\relax\ifmmode{\text{MeV}}              
                     \else{MeV{ }}\fi}                    
\def\MSbar{\relax\ifmmode\overline                        
            {\rm MS}\else{$\overline{\rm MS}${ }}\fi}     
\def\as{\relax\ifmmode \alpha_s\else{$ \alpha_s${ }}\fi}  
\def\abar{\relax\ifmmode{\bar{a}}\else{$\bar{a}${ }}\fi}  
  \def\ie{\hbox{\it i.e.}{ }} 
   \def\eg{\hbox{\it e.g.}{ }}  
\newcommand{\Tf}{ T_f }
\newcommand{\dRRNA}{\frac{d_F^{abcd}d_F^{abcd}}{N_A}}
\newcommand{\dRANA}{\frac{d_F^{abcd}d_A^{abcd}}{N_A}}
\newcommand{\dAANA}{\frac{d_A^{abcd}d_A^{abcd}}{N_A}}

\newcommand{\tr}{T_\text{R}}
\newcommand{\Ngl}{n_{\tilde{g}}}
\newcommand{\Nf}{n_{f}}
\usepackage{color}
\newcommand{\red}[1]{{\color{red} #1}}
\newcommand{\blue}[1]{{\color{blue} #1}}
\definecolor{green}{rgb}{0.133,0.56,0}

\definecolor{DarkGreen}{rgb}{0.04,0.5,0.1}
\newcommand{\DarkGreen}[1]{{\color{DarkGreen} #1}}

\def\1{\hbox{{1}\kern-.25em\hbox{l}}}

\title{On a realization of $\{\beta\}$-expansion in QCD}

  \author[a]{S.~V.~Mikhailov}

   \affiliation[a]{Bogoliubov Laboratory of Theoretical Physics, JINR,
                141980 Dubna, Russia}
\emailAdd{mikhs@theor.jinr.ru}
\abstract{%
 We suggest a simple algebraic  approach to fix the elements of the $\{ \beta \}$-expansion for
 renormalization group invariant quantities, which uses additional degrees of freedom.
The approach is discussed in detail for N$^2$LO calculations in QCD with the MSSM gluino -- an add\-itional degree of freedom.
We derive the formulae of the $\{ \beta \}$-expansion for
the nonsinglet Adler $D$-function and Bjorken polarized sum rules in
the actual N$^3$LO within this quantum field theory scheme with the MSSM gluino and
the scheme with the second additional degree of freedom.
We discuss the properties of the $\{ \beta \}$-expansion for higher orders considering
the N$^4$LO as an example.
}
\begin{document}
\maketitle
\section{Introduction}
\label{intro}
The knowledge of the detailed structure of QCD perturbation expansions is rather
important for a variety of tasks of which the renormalization group optimization of the
series is the best known.
The detailed structure looks as a double series (or a matrix representation) rather than a usual series
\cite{Mikhailov:2004iq,Kataev:2010du} even in the case of expansion of
`physical' quantities (having no anomalous dimension).
We shall explore this structure  for the QCD renormalization group  invariant
(RGI) one-scale  dependent quantities as well as
elaborate an algebraic approach to fix their elements within the $\{\beta\}$-expansion \cite{Mikhailov:2004iq}.
In this sense  our paper continues the  investigations of the perturbation
expansions in \cite{Kataev:2010du,Kataev:2014jba,Kataev:2016aib}.
Now let us introduce the appropriate physically important quantities whose
perturbation expansions are the most advanced.
We take as patterns of the RGI quantities the phenomenological important
Bjorken polarized sum rules $S^\text{Bjp}(Q^2,\mu^2)$,
\ba
&&S^\text{Bjp}(Q^2)
=\frac{g_A}{6}~\left[
C^\text{Bjp}_{\rm NS}(Q^2/\mu^2,a_s(\mu^2)) +
\left ( \sum_{i} q_i \right )C^\text{Bjp}_{\rm S}(Q^2/\mu^2,a_s(\mu^2))\right]\,,
\ea
and the Adler function $D(Q^2,\mu^2)$,
\ba
\label{DA}
\!\!\!\!D^{\rm EM}\left(\frac{Q^2}{\mu^2},a_s(\mu^2)\right)\!\! =\!\! \left(  \sum_i q_i^2
\right)\!\!
d_R D_{\rm NS}\left(\frac{Q^2}{\mu^2},a_s(\mu^2)\right)
\!+\! \left(\sum_i q_i \right)^2\!\! d_R D_{\rm S}\left(\frac{Q^2}{\mu^2},a_s(\mu^2)\right),
\ea
where $q_i$ is the electric charge of the quark, $g_A$ -- nucleon
axial charge, $d_R$ -- the dimension of the quark color representation.
Perturbative expression for the nonsinglet (NS) 
coefficient functions of both the quantities at the renormalization scale $\mu^2=Q^2$ can be written down as
\begin{eqnarray} \label{eq:PT-D-C}
  D_{\rm NS}(a_s(\mu^2)) = 1 + \sum_{n\geq1} a^n_s(\mu^2)~d_n, ~~~C^\text{Bjp}_{\rm NS}(a_s(\mu^2)) =    1 + \sum_{l\geq 1}a_s^l(\mu^2) \ c_l;
 \end{eqnarray}
 $a_s =\alpha_s/(4\pi)$, they are calculable in the \MSbar-scheme
 and were obtained in order of $O(a_s^4)$ in \cite{Baikov:2010je}.
 We use here only the NS parts of these quantities,
 $D=D_\text{NS},~C^\text{Bjp}=C^\text{Bjp}_\text{NS}$, omitting
the  corresponding notation further in the text.
The  perturbation  coefficients $d_n ~(c_n)$ in Eqs.(\ref{eq:PT-D-C})  are the combinations
of only the color coefficients.

Now, recall the structure of these perturbation coefficients.
The $\{\beta\}$-expansion representation introduced in \cite{Mikhailov:2004iq}
 prescribes to decompose  $d_n$or/and $c_n$ or any other of RGI quantity in the following way:
\begin{subequations}
\label{eq:d_beta}
\begin{eqnarray}
\label{eq:d_1}
d_1&=&d_1[0]\, , \\
d_2&=&\! \beta_0\,d_2[1]
  + d_2[0]\, ,\label{eq:d_2}\\
  d_3
&=&\!
  \beta_0^2\,d_3[2]
  + \beta_1\,d_3[0,1]
  +       \beta_0 \,  d_3[1]
  + d_3[0]\, ,\label{eq:d_3} \\
  d_4
   &=&\! \beta_0^3\, d_4[3]
     + \beta_1\,\beta_0\,d_4[1,1]
     + \beta_2\, d_4[0,0,1]
     + \beta_0^2\,d_4[2]
     + \beta_1  d_4[0,1]
     + \beta_0\,d_4[1] \nonumber \\
   && \phantom{\beta_0^3\, d_4[3]+ \beta_1\,\beta_0\,d_4[1,1]+ \beta_0^2\,}~+d_4[0]\,,
       \label{eq:d_4} \\
   &\vdots& \nonumber \\
 d_{n}
   &=&\!\!\beta_0^{n-1}\, d_{n}[n\!-\!1]+ \cdots + d_n[0]\,,
\label{eq:d_n}
\end{eqnarray}
\end{subequations}
where  $\beta_i$ are the coefficients of the QCD $\beta$-function
\begin{equation}
\label{eq:beta}
\mu^2\frac{d a_s(\mu^2)}{d \mu^2}=
\beta(a_s)=-a_s^{2}(\mu^2) \sum_{i\geq 1} \beta_{i-1} a_s^{i-1}(\mu^2)\,,
\end{equation}
and the explicit expressions for $\beta_{0-3}$ are presented in Appendix \ref{App:A}.
The notation $i_0, i_1,\ldots$ of the arguments of $d_n[i_0,i_1,\ldots]$ denotes
the powers of accompanying $\beta_0, \beta_1,\ldots$.
The elements $d_n[\cdot]$ of the decomposition do not depend on the number
of active quarks $n_f$ at least up to the actual order $O(a_s^3)$ (n=3).
This important property of the elements $d_n[\cdot]$ will be discussed at the beginning
of Sec.\ref{sec:3} and in Sec.\ref{sec:4}.
The decompositions in Eqs.(\ref{eq:d_beta}) should contain the complete knowledge
about strong charge renormalization by means of using there all of the possible
$\beta_i$-terms \cite{Kataev:2010du,Kataev:2014jba}
(see the beginning of Sec.\ref{sec:4} for details).
This kind of the expansion is the essential part of the procedures for the optimization of
perturbation series, \eg, the decomposition (\ref{eq:d_2})
 was the starting point of the well-known BLM prescription
\cite{Brodsky:1982gc} in NLO; for further high order development
 see \cite{Mikhailov:2004iq,Kataev:2014jba,Mojaza:2012mf,Ma:2015dxa}
 and the references cited therein.

At NLO of QCD the decomposition in (\ref{eq:d_2}) looks evident because the term proportional
to $\frac{4}3 T_\text{R} n_f$
unambiguously marks the contribution of the term proportional to $  \beta_0
   = \frac{11}{3} C_A - \frac{4}{3} T_R n_f$
 in $d_2$, see the discussion in Sec.3B in
\cite{Kataev:2014jba} and the result in Eq.(\ref{D-21}) in Appendix \ref{App:A}.
 How to fix the elements of the decomposition in higher orders?
 The consideration  of color coefficients content of  the $d_n$ is not enough for this,
so one need to find additional conditions.
To solve the problem, we introduce additional degrees of freedom (d.o.f.),
 new fields that interact following the universal gauge principle and
  enter only in intrinsic loops.
Using the fermions in the adjoint representation (MSSM light gluino) as an additional
d.o.f., we formulate a simple algebraic scheme to obtain the elements of the $\{\beta\}$-expansion and demonstrate the results in N$^2$LO,
Eq.(\ref{eq:d_3}), in the following Sections~\ref{sec:example} and \ref{sec:2.2}.
Moreover, based on the Crewther relation \cite{CR} we derive the relation between
$C$ and $D$ in Sec.~\ref{sec:2.3}.
This algebraic scheme is well algorithmized  and appropriate to apply to high loop results.
It is applied to the N$^3$LO expansion in  Eq.(\ref{eq:d_4}) in Sec.\ref{sec:3},
which fixes completely the elements of expansion.
The required expressions for $d_4,~c_4$ with the additional d.o.f. are expected to be calculated
in future.
In Sec.\ref{sec:4},  we discuss the  general structure and properties
of $\{\beta\}$-expansion for higher orders considering the N$^4$LO
as an example.
The algebraic scheme fixes the expansion elements for this case too.
   Our main results are presented in Conclusion.

\section{Algebraic approach for the $\{\beta\}$-expansion in N$^2$LO}
\label{sec:2}
\subsection{A simple illustration}
 \label{sec:example}
Let us consider the task to fix the decomposition elements in Eq.(\ref{eq:d_beta}) algebraically,
taking Eq.(\ref{eq:d_3}) as an example.
The ``renormalon'' term $d_3[2]$ (or $d_n[n-1]$ for any $n$) at the maximum power of $\beta_0$ can be identified by the maximum power of
$\frac{4}3 T_\text{R} n_f$ (here it is proportional to
$C_\text{F}~\left(\frac{4}3 T_\text{R} n_f \right)^2$~),
or even calculated independently, see \cite{Broadhurst:1993ru}.
The corresponding residual in the RHS of Eq.(\ref{eq:d_3}) contains 5 Casimir coefficients $C^3_\text{F},~C^2_\text{F}C_\text{A},  ~C_\text{F}C^2_\text{A},~C^2_\text{F}T_\text{R} n_f,~~C_\text{F}C_\text{A}T_\text{R} n_f $
that are distributed among three terms $d_3[\cdot]$ there.
Finally we need to obtain these three unknown elements $d_3[0,1], d_3[1], d_3[0]$ in the RHS of Eq.(\ref{eq:d_3X})
\begin{eqnarray}
 \label{eq:d_3X}
  \bar{d}_3(x)\equiv d_3(x)- \beta_0^2(x)\,d_3[2]
&=&\!
  \beta_1(x)\,d_3[0,1]
  +       \beta_0(x) \,  d_3[1]
  + d_3[0]\,,
\end{eqnarray}
where we put variable $x=\frac{4}{3} T_R n_f$.
Taking Eq.(\ref{eq:d_3X}) at any three different values of $x,~(x_1, x_2, x_3)=X$ and compiling
the coupled system of linear equations we can obtain the unique solution of this system under the
evident condition that the corresponding determinant $\Delta_3$,
\begin{eqnarray}
 \label{eq:delta3}
\!\!\!\! \Delta_3(X)= \left(\beta_0(x_2)-\beta_0(x_1) \right)\left(\beta_1(x_1)-\beta_1(x_0) \right)-
          \left(\beta_0(x_1)-\beta_0(x_0) \right)\left(\beta_1(x_2)-\beta_1(x_1) \right),
\end{eqnarray}
is not zero.
The opposite condition $\Delta_3=0$ unambiguously means that the functions $\beta_0(x),\beta_1(x)$
are linear in  $x$,
\be
\frac{\beta_0(x_2)-\beta_0(x_1)}{\beta_0(x_1)-\beta_0(x_0)}=
\frac{\beta_1(x_2)-\beta_1(x_1)}{\beta_1(x_1)-\beta_1(x_0)}\,, \nonumber
\ee
this is just the case of QCD with only quark degrees of freedom
(see the explicit expressions in Eq.(\ref{eq:beta0-3}) ).
Due to this reason one cannot untangle contributions from $\beta_0$ and $\beta_1$ in N$^2$LO
without an additional constraint (see the discussion in \cite{Kataev:2014jba}).
In the case of an additional degree of freedom that contributes to both sides of Eq.(\ref{eq:d_3X}),
\ie, to the coefficient $d_3$ and to $\beta_0, \beta_1$,
one can obtain the unique solution.
 The goal of this note is to elaborate an algebraic scheme to obtain the decompositions in Eqs.(\ref{eq:d_beta})
 using additional d.o.f.
like $n_{\tilde{g}}$ -- the number of MSSM gluino (we use $\Ds y=\frac{4}3\frac{C_\text{A}}2 n_{\tilde{g}}$) and, may be, other fields that interact following the universal gauge principle
and appear only in intrinsic loops.
The net effect of this field will be parameterized by means of the parameter $z$.
Further, we shall suggest that the coefficients of perturbation expansion,
like $d_n~(c_n)$ in the LHS of (\ref{eq:d_beta}),
 as well as the coefficients of the $\beta$-function in the RHS of (\ref{eq:d_beta}) are calculated within the \MSbar scheme
and are  known  functions on the arguments $x,y,z$.
To be more exact we consider the Adler function $D(x,y)$ \cite{Chetyrkin:1996ez}
as well as the $\beta$-coefficients
$\beta_0(x,y), \beta_1(x,y)$ presented in Appendix \ref{App:B} as functions on both the quark $(x)$ and the MSSM gluinos $(y)$ d.o.f.
In this notation $\beta_i(x,0)=\beta_i(x),~d_n(x, 0)=d_n(x), \ldots$.
The results for the decomposition presented below are valid also for the coefficient function
$C^\text{Bjp}(x,y)$ up to the replacement of
the notation and for any RGI one-scale quantities.

\subsection{The formalism of decomposition for D-function}
\label{sec:2.2}
To simplify the system of equations (SE) based on Eq.(\ref{eq:d_3XY}) (the extended
by the $y$ d.o.f. Eq.(\ref{eq:d_3X})),
\begin{eqnarray}
 \label{eq:d_3XY}
  \bar{d}_3(x,y)\equiv d_3(x,y)- \beta_0^2(x,y)\,d_3[2]
&=&\!
  \beta_1(x,y)\,d_3[0,1]
  +       \beta_0(x,y) \,  d_3[1]
  + d_3[0]\,,
\end{eqnarray}
we take for the components of
$X$: $x_0, x_1, (x_{01},y_{01})$, the special values -- the roots of equations
\be
\label{eq:3X}
 \beta_0(x_0)=0,~ \beta_1(x_1)=0,~\{\beta_0(x_{01},y_{01})=0,~\beta_1(x_{01},y_{01})=0\}.
  \ee
For this $X_3$ the SE$_3$ looks like
\ba
 \label{eq:SE3}
\left\{
 \begin{array}{rcl}
 d_3(x_{01},y_{01})&=&d_3[0] \\
  d_3(x_{0},0)&=&\beta_1(x_0)d_3[0,1]+d_3[0] \\
   \bar{d}_3(x_{1},0)&=&\beta_0(x_1)d_3[1]+d_3[0]\,.
 \end{array}
\right.
 \ea
Now the value of the determinant $\Delta_3(X_3)\!=\! \beta_0(x_1)\beta_1(x_0)\!=$$\Ds\!-\frac{{\rm C_A^2}
\left(7C_\text{A}+11C_\text{F} \right)^2}{\left(5C_\text{A}+3C_\text{F} \right)}$ $\! \neq \!0$ that follows
 from Eq.(\ref{eq:delta3}) or, can be obtained from the SE$_3$ (\ref{eq:SE3}) directly.
Therefore the unique solution of the SE$_3$  is
\begin{subequations}
 \label{eq:d_3sol}
\begin{eqnarray}
d_3[0]&=& d_3(x_{01},y_{01})\,, \\
 d_3[0,1]&=&\big(d_3(x_0) -d_3(x_{01},y_{01})\big)/\beta_1(x_0)\,,\\
  d_3[1]&=&\big(\bar{d}_3(x_1) -d_3(x_{01},y_{01})\big)/\beta_0(x_1)\,.
  \end{eqnarray}
\end{subequations}
 These values were obtained first in \cite{Mikhailov:2004iq} using another trick;
here they are presented explicitly in Eq.(\ref{eq:d1-4}) in Appendix \ref{App:A}.
\subsection{How to obtain the $\{\beta\}$-expansion for  $C^\text{Bjp}$ from one for $D$}
 \label{sec:2.3}
To relate the already known structure of $d_3$ (the solutions in (\ref{eq:d_3sol})) to the corresponding
$\{\beta\}$-expansion of $c_3$, we use the generalized Crewther relation (CR) \cite{Kataev:2010du,Broadhurst:1993ru},
\begin{equation}
\label{MCre}
 D(a_s) \cdot C^\text{Bjp}(a_s) = \1 + \beta(a_s)\,\cdot K(a_s)\,,
\end{equation}
where $K(a_s)=\sum_{n=1}a_s^{n-1}K_n$ is a polynomial in $a_s$.
In the case of the $\beta$-function having  identically zero coefficients
$\beta_i=0$,  the generalized CR (\ref{MCre}) returns to its initial form
 \cite{CR} with only $\1$ in its RHS that expresses the unbroken conformal symmetry.
The later condition relates the $d_n[0],~c_n[0]$ elements in every order (see definition (4.1) and
Eq.(4.2) in \cite{Kataev:2014jba})
\ba
c_n[0] &=& -d_n[0] - \sum_{l=1}^{n-1} d_l[0] c_{n-l}[0]\,. \label{eq:CI-PT0}
\ea
The explicit closed solution of the relation (\ref{eq:CI-PT0}) with respect to $c_k[0]$ is
\ba \label{eq:CI-PT0n}
c_k[0]= (-)^k \text{det}[D^{(k)}_0] \equiv (-)^k \left|
 \begin{array}{lllccc}
                 d_1 & 1    & 0   &            &\ldots & 0 \\
                 d_2 & d_1  & 1   &            &\ldots & 0 \\
                 d_3 & d_2  & d_1 &            &\ldots& 0  \\
              \ldots &      &     &            &\ldots& 0  \\
              d_{k-1}&\ldots&\ldots&            & d_1  & 1  \\
                 d_k & d_{k-1} & d_{k-2}&\ldots&d_2 & d_1   \\
               \end{array}
             \right|\,,
\ea
here $D^{(k)}_0$ -- matrix, which consists of $d_i\equiv d_i[0]$ elements.
 The general relation (\ref{eq:CI-PT0n}) can be also treated as a prediction
for $C^\text{Bjp}$ by means of $D$ that is based on the  $\{\beta\}$-expansion and CR.
In the third order of $a_s$
the knowledge of the element $c_3[0] = - d_3[0]+2d_1 d_2[0]-(d_1)^3 $,
followed from (\ref{eq:CI-PT0}), allows us to fix  all the other elements
of the expansion in this order, \ie to disentangle the contributions from $c_3[1]\beta_0$ and $c_3[0,1]\beta_1$,
which were discussed in detail in Sec.IV~B in \cite{Kataev:2014jba}.

From another side one can use in the RHS of Eq.(\ref{MCre}) the second term proportional to $\beta(a_s)$
that expresses the conformal symmetry-breaking.
This leads to the series of relations \cite{Kataev:2010du,Kataev:2014jba}
for the elements of the different orders $n$ at $\beta_{n-1}$, \eg,
\begin{eqnarray}
  \label{eq:k_n1-d_n1}
\!\!\!\!\!\! d_2[1] + c_2[1]\!=\!d_3[0,1] + c_3[0,1]\!=\!\! \ldots  
\!\!=\! d_n[\underbrace{0,0,\ldots, 1}_{n-1}] + c_n[\underbrace{0,0,\ldots, 1}_{n-1}]\!=\!\!
3C_\text{F}\! \left(\frac{7}2-4\zeta_3 \right)\!.
 \end{eqnarray}
In the third order this gives the equation $c_3[0,1]=d_2[1]+c_2[1]-d_3[0,1]$,
which fixes $c_3[0,1]$ and also admits restoration of two other terms $c_3[1],~c_3[0]$.

Both of the ways provide the same results for the elements
$c_3[0,1], c_3[1]$, $c_3[0]$ of the $\{\beta\}$-expansion \cite{Kataev:2014jba}.
As a byproduct of the procedure we predicted the $c_3$ of $C^\text{Bjp}$,
see Eq.(4.11) in \cite{Kataev:2014jba},
if the $\{\beta \}$-expansion for $d_3$ of $D$ is already known and vice versa.

\section{The $\{\beta\}$-expansion for Bjorken polarized SR and D-function in N$^3$LO}
\label{sec:3}
In the 5 loop case, $d_4(x)$ was first obtained in \cite{Baikov:2008jh} as the polynomial
with numerical coefficients, then all the color coefficients
 in decomposition for $d_4(x)$ and $c_4(x)$ were presented in \cite{Baikov:2010je}.
 Following the $\{\beta\}$-expansion we propose for these coefficients the decomposition,
\begin{eqnarray}
 \label{eq:d_4XY}
\!\!\!\!\!\!\!\!\! \bar{d}_4(x,y)&\equiv & d_4(x,y)- \beta_0^3\, d_4[3] \nonumber \\
  & =&
     \beta_1\,\beta_0\,d_4[1,1]
     + \beta_2\, d_4[0,0,1]
     + \beta_0^2\,d_4[2]
     + \beta_1  d_4[0,1]
     + \beta_0\,d_4[1]+d_4[0]\,.
\end{eqnarray}
Here one has six unknown elements
$d_4[1,1],\, d_4[0,0,1],\, d_4[2],\, d_4[0,1],\, d_4[1],\, d_4[0]$,
while the seventh element $d_4[3]$ can be directly identified.
The ten Casimirs (here $T_f \equiv T_\text{R}n_f$)
$
C_F^4\,,
 C_F^3 \Tf \,,
  C_F^2 \Tf^2 \,,C_F   \Tf^3 \,$,
  $
\,  C_F^3 C_A\,,
\,  C_F^2 \Tf  C_A,
C_F \Tf^2 C_A  \,,
\,  C_F^2 C_A^2\,,
    C_F  \Tf C_A^2\,,
\,  C_F C_A^3
$
are dis\-trib\-uted among all of the $d_4[\cdot]$ elements,
while the abelian elements of the box subgraphs with four gluon legs,
related to color coefficients
$
n_f\, d_F^{a b c d}d_F^{a b c d}/d_R
$ (quark box inside),
$ \Ds
d_F^{a b c d} d_A^{a b c d}/d_R\,
$ (gluon box inside),
 enter into $d_4[0]$.
These terms do not contribute to the renormalization of the
charge\footnote{I thank A.~Grozin for clarifying this subject} $a_s$,
see also the discussion of the subject in \cite{Cvetic:2016rot}.
Although the corresponding 5-loop diagrams contain these one-loop boxes,
further contraction of the subgraphs
(see the discussion in \cite{Kataev:2016aib}) do not contribute to $\beta_0$.
 Due to this reason $d_4[0]$ get $(x,y)$-dependence $d_4[0] \to d_4[0](x,y)$.
 We decompose it as $d_4[0](x,y)=\tilde{d}_4[0]+\delta d_4(x,y)$,
 where the $(x,y)$-dependent part $\delta d(x,y)$ is well recognized,
 while $\delta d_4(x,0)(\delta c_4(x,0))$ is already known from the result in \cite{Baikov:2010je} (see Eq.(\ref{eq:A6}) in Appendix~\ref{App:A}).
 Therefore, the $n_f~(n_{\tilde{g}} )$-dependence becomes partly separated from the charge
 renormalization for the first time in N$^3$LO.
\subsection{Decomposition with 2 degrees of freedom $x, y$}
 \label{sec:3.1}
To obtain these $d_4[\cdot]$ elements,
one can take six points $X$ in the plane $(x,y)$.
Then one takes the set $X$ as the arguments of $d_4(x,y)$ and $\beta_i(x,y)$
and  compiles the system of linear equations (SE$_6$), based on Eq.(\ref{eq:d_4XY}),
with respect to these six unknown elements of the $\beta$-expansion.
Again, to simplify the calculation, we take for these six components of
$X_6$: $x_0, x_1, x_2$, $ (x_{01},y_{01})$, $(x_{02},y_{02})$, $(x_{12},y_{12})$  the roots of the equations and
the systems of equations
\begin{subequations}
\label{eq:6-2-cond}
\ba
 \!\!&&\!\!\!\!\!\!\beta_0(x_0,0)=0, \beta_1(x_1,0)=0, \beta_2(x_{2m(p)},0)=0,\{\beta_0(x_{01},y_{01})=0, \beta_1(x_{01},y_{01})=0\}, \label{eq:6-2-cond-a}\\
\!\! &&\!\!\!\!\!\! \{\beta_0(x_{02},y_{02})=0,\beta_2(x_{02},y_{02})=0\},
  \{\beta_1(x_{12},y_{12})=0,\beta_2(x_{12},y_{12})=0\}.\label{eq:6-2-cond-b}
  \ea
  \end{subequations}
 We shall supply the solutions in Eqs.(\ref{eq:6-2-cond-b}) the subscripts:
 $\Ds x_{02m(p)}, y_{02m(p)}, x_{12m(p)}, y_{12m(p)}$,
 to separate the different roots $m (-), p(+)$ of quadratic
 equations for the cases where $\beta_2(x,y)$ are involved (see the expression in (\ref{eq:beta-b2})).
  The determinant $\Delta_6(X_6)$ of the corresponding SE$_6$ is
\begin{eqnarray}
\!\!\!\!\!\!\!\!\! \Delta_6(X_6)\! &=&\! \beta_0(x_1)\beta_0(x_{2m})\beta_1(x_{2m}) \Re{\beta_0(x_{12m},y_{12m})}
 \left[\beta_0(x_1)- \Re \beta_0(x_{12m},y_{12m}) \right] \delta_6, \\
\!\!\!\!\!\!\!\!\! \delta_6\!&=&\! \left[\beta_1(x_{02m},y_{02m} )\beta_2(x_{0})- \beta_1(x_{02m},y_{02m} )\beta_2(x_{01},y_{01})+
 \beta_1(x_{0}) \beta_2(x_{01},y_{01}) \right]\,.
 \end{eqnarray}
 This value $\Delta_6(X_6) \neq 0$, the solution of this SE$_6$ exists and unique,
 and can be obtained like the solution of Eq.(\ref{eq:d_3sol}) for the N$^2$LO in Sec.\ref{sec:2.2}.
 Therefore, to derive the $\{\beta\}$-expansion for $d_4$,  it is
 \textit{enough to obtain one at an additional single} d.o.f. $y$, $d_4 \to d_4(x,y)$
 together with the coefficients $\beta_{0,1,2}(x,y)$ (see Appendix \ref{App:B}).

We present the solutions of SE$_6$ for a number of elements in the explicit form,
taking the notation for the arguments $(x_{ij},y_{ij})$ and the function $Y_4$ for shortness,
\be \label{def:rY}
r_{ij}=(x_{ij},y_{ij}),~~Y_4(X_6)\equiv \bar{d}_4(X_6)-\delta d_4(X_6),
\ee
\begin{subequations}
 \label{eq:d_4Sol}
\begin{eqnarray}
\!\!\!\!\!\!\!\!\!\!\!\!\!d_4[0,0,1]&=&\!\!\left[Y_4(r_{01})(\beta_1(x_0) -\beta_1(r_{02}))-Y_4(r_{02})\beta_1(x_0)+Y_4(x_0)\beta_1(r_{02})\right]/\delta_6,\\
\!\!\!\!\!\!\!\!\!\!\!\!\!d_4[0,1]&=&\!\!\left[Y_4(r_{02})(\beta_2(x_0) -\beta_2(r_{01}))-Y_4(r_{01})\beta_2(x_0)+Y_4(x_0)\beta_2(r_{01})\right]/\delta_6, \\
\!\!\!\!\!\!\!\!\!\!\!\!\! \tilde{d}_4[0]&=&\!\!\left[Y_4(r_{02})\beta_2(r_{01})\beta_1(x_0)+Y_4(r_{01})\beta_1(r_{02})
\beta_2(x_0)-Y_4(x_0)\beta_2(r_{01})\beta_1(r_{02})\right]/\delta_6.
 \end{eqnarray}
  \end{subequations}
  Just these elements will be used for the relation with  similar elements in $C^\text{Bjp}$.

Of course, one can take another set $X'_6$ and construct the corresponding SE$'_6$.
In any case the solution for the elements $d_4[\cdot]$ should be the same.
The usage of the roots in  Eqs.(\ref{eq:6-2-cond}) to construct $X_6$ leads to the
simplification of the final SE.

\subsection{Relations between the elements of $D$ and $C^\text{Bjp}$}
 \label{sec:3.2}
Suppose that the coefficient functions for the Adler D-function $d_4(x,y)$ and the Bjorken SR $c_4(x,y)$
are  known.
Then, based on two terms in the RHS of the Crewther relation, Eq.(\ref{MCre}),
one can obtain for the sum of these functions $d_4(x,y)+c_4(x,y)$ and their elements (see \cite{Kataev:2010du})
a series of the relations.
In part, one can obtain from (\ref{eq:CI-PT0}) the sum of ``zero'' elements,
\begin{equation}
 d_4[0](x,y)+c_4[0](x,y)= \tilde{d}_4[0]+\tilde{c}_4[0]= 2d_1 d_3[0]-3d_1^2d_2[0]+d_2[0]^2+d_1^4,
 \end{equation}
the term $\delta d_4(x,y)$ is cancel in the sum $d_4[0]+c_4[0]$.
Generally speaking, for $n$-order case

\ba
c_n[0] + d_n[0] = - \sum_{l=1}^{n-1} d_{n-l}[0] (-)^l \text{det}[D^{(l)}_0]=  (-)^n \left|
\begin{array}{lllccc}
                 d_1 & 1    & 0   &            &\ldots & 0 \\
                 d_2 & d_1  & 1   &            &\ldots & 0 \\
                 d_3 & d_2  & d_1 &            &\ldots& 0  \\
              \ldots &      &     &            &\ldots& 0  \\
              d_{k-1}&\ldots&\ldots&            & d_1  & 1  \\
                 \bm{0} & d_{k-1} & d_{k-2}&\ldots&d_2 & d_1   \\
               \end{array}
             \right|\,, \label{eq:CI-PTsum}
\ea
where $D^{(l)}_0$ is defined in (\ref{eq:CI-PT0n}).
The LHS of Eq.(\ref{eq:CI-PTsum}) is of $n$-order, while its RHS dependents on the $d_l[0]$ elements of less orders $ l \leqslant n-1$, therefore this equation can serve a good check
for next order results.
The others relations are:
 \begin{subequations}
  \label{eq:CD}
  \begin{eqnarray}
\!\!\!\!\!\!\!\! \left(d_4 +c_4 \right)(x_{01}, y_{01})
\!\! &=&\!\! \beta_2(x_{01}, y_{01})\left(\underline{d_4[0,0,1]+c_4[0,0,1]} \right)+\underline{\underline{d_4[0]+c_4[0]}}= \label{eq:CDa} \\
\!\! &=&\!\! \beta_2(x_{01}, y_{01})\underline{3 C_\text{F}\!\!\left(\frac{7}{2} -4\zeta_3 \right)}\!\!+
  \!\! \left(\underline{\underline{2d_1 d_3[0]-3d_1^2d_2[0]+ d_2[0]^2+d_1^4}} \right)\label{eq:CD01} \\
  \!\! &=&\!\!\beta_2(x_{01}, y_{01})\underline{3 C_\text{F}\!\!\left(\frac{7}{2} -4\zeta_3 \right)}\!\!+
  \!\! \underline{\underline{(3 C_\text{F})^2\left[ \left(\frac{175}6 -144 \zeta_3 \right) {\rm C_A^2} +
                        44{\rm C_F C_A} - \frac{37}4{\rm C_F^2}\right]}} \nonumber
   \end{eqnarray}
    \begin{eqnarray}
     \left(d_4 +c_4 \right)\left(x_{02}, y_{02}\right)
   &=&\beta_1(x_{02}, y_{02})\left(\underline{d_4[0,1]+c_4[0,1]} \right)+d_4[0]+c_4[0]= \label{eq:CDc} \\
   &=&\beta_1(x_{02}, y_{02})\underline{3C_\text{F}} \Big[\underline{C_\text{A}\left(\frac{47}9-\frac{16}3\zeta_3 \right) +C_\text{F}
   \left(-\frac{397}{18}-\frac{136}3\zeta_3+80\zeta_5\right)}+  \nonumber \\
   &&\underline{3C_\text{F} \left(\frac{40}3 -12\zeta_3 \right)}\Big] +
     (3 C_\text{F})^2\left[ \left(\frac{175}6 -144 \zeta_3 \right) {\rm C_A^2} +44{\rm C_F C_A}  -
\right. \nonumber \\
    && \left. \phantom{3C_\text{F} \left(\frac{40}3 -12\zeta_3 \right)\Big]+ (3 C_\text{F})^2\left(\frac{175}6 -144 \zeta_3 \right)}
    ~ \frac{37}4{\rm C_F^2}\right] \,, \label{eq:CD02}
   \end{eqnarray}
    \begin{eqnarray}
\!\!\!\!\!\!\!\!\!\!     \left(d_4 +c_4 \right)\left(x_{0}\right)
   &=&\!\! \beta_2(x_{0})\left(d_4[0,0,1]+c_4[0,0,1]\right)+\beta_1(x_{0})\left(d_4[0,1]+c_4[0,1] \right)+ \nonumber \\
     &&d_4[0]+c_4[0]= \label{eq:CDe} \\
    &=&3{\rm  C_F}\Bigg[-\frac{111}{4}{\rm  C_F^3}+{\rm C_A C_F^2} \left(-\frac{1661}{36}+
\frac{2618}{3}\zeta_3- 880\zeta_5\right)+ \nonumber \\
\!\!\!\!\!\!&&\!\! {\rm C_A^2 C_F}  \left(-\frac{3337}{18}+\frac{896}{3} \zeta_3-
   35 16 \zeta_5\right)+
  {\rm  C_A^3 } \left(-\frac{28931}{144}+\frac{1351}{6}
   \zeta_3\right)\Bigg].
\end{eqnarray}
 \end{subequations}
The RHSs of (\ref{eq:CDa}, \ref{eq:CDc}, \ref{eq:CDe}) are presented by mean of the already
known results for $d_3$ and $c_3$ -- (\ref{eq:CD01}, \ref{eq:CD02}),
see \cite{Kataev:2014jba} and Appendix \ref{App:A} here.
The last Eq.(\ref{eq:CDe}), suggested in \cite{Kataev:2010du}, has already been verified and is put here
for illustration and comparison with the two previous equations.
Let us conclude,

(i) For the values of $d_4 +c_4$ on $(x_{01}, y_{01})$ and $(x_{02}, y_{02})$ Eqs.(\ref{eq:CD})
provide the simple check -- the RHS of (\ref{eq:CD01}) and (\ref{eq:CD02}), respectively.

(ii)The equalities of the underlined terms in  Eqs.(\ref{eq:CD}) are realized independently
following Eqs.(\ref{eq:k_n1-d_n1})  and (\ref{eq:CI-PT0}).
Therefore, the second equalities in (\ref{eq:CDa}) and (\ref{eq:CDc}) allow one to get
$d_4[0,0,1], d_4[0,1], d_4[0]$ through their partners $c[\cdot]$ and vice versa using the solutions
in Eq.(\ref{eq:d_4Sol}), see Eqs.(30),
(31)\footnote{There in the RHS of Eq.(31) was missed the term $ +(-47/48+\zeta_3)C_\text{F}C_\text{A} $}
 in \cite{Kataev:2010du}.
But, one cannot restore all the elements of $C$ in N$^3$LO based only on CR
and the known $D$ opposite to the case of N$^2$LO, see Sec.\ref{sec:2.3}.

Let us mention thereupon an alternative approach to fix $d_4[\cdot],~c_4[\cdot]$
without additional d.o.f.,
which was suggested in \cite{Kataev:2016aib} and was inspired by the structure of the
RHS of CR (\ref{MCre}).
The idea is based on the specific proposition that the perturbation series for $D$ and $C^\text{Bjp}$
can be expanded in powers $( \beta(a_s)/a_s )^n $ similar to that had been proposed for the ``conformal symmetry braking term'' $ \beta(a_s) K(a_s)$ in the  RHS of CR, see the presentation in Eq.(6) in \cite{Kataev:2010du}.
The results for the elements obtained within this approach differ from ours.
\subsection{What can we get at 3 degrees of freedom $x, y, z$}
 \label{sec:3.3}
Let us imagine that we have an additional third ``intrinsic'' d.o.f. that
manifests itself as the parameter $z$.
In this case $d_n=d_n(x, y, z), ~\beta_i=\beta_i(x, y, z)$; therefore, one can use the points
in $(x, y, z)$ space to construct the set $X$:
\begin{equation}
 \label{eq:012}
\{\beta_0(x_{012}, y_{012}, z_{012})=0,\beta_1(x_{012}, y_{012}, z_{012})=0, \beta_2(x_{012}, y_{012}, z_{012})=0\}
\end{equation}
instead of the later constraint in Eq.(\ref{eq:6-2-cond-b}) ( if the solution of SE (\ref{eq:012}) exists).
Let us call this solution $r_{012}=(x_{012}, y_{012}, z_{012})$ and
$\beta_i(x)=\beta_i(x,0,0),~\beta_i(x,y)=\beta_i(x,y,0)$,~$d_n(x)=d_n(x, 0,0),~d_n(x,y)=d_n(x, y,0), \ldots$ for shortness.
From Eq.(\ref{eq:d_4XY}) and Eq.(\ref{eq:012}) it immediately follows that
\begin{eqnarray}
                   d_4(r_{012})&=& \tilde{d}_4[0]+ \delta d(r_{012}), \label{eq:XYZd40}\\
     c_4(r_{012})+d_4(r_{012}) &=&\tilde{d}_4[0]+\tilde{c}_4[0]= 2d_1 d_3[0]-3d_1^2d_2[0]+d_2[0]^2+d_1^4. \label{eq:c4+d4}
\end{eqnarray}
Equation (\ref{eq:c4+d4}) provides an independent test for the $c_4,d_4$ results.
In the case of constraint (\ref{eq:012}), the procedure for obtaining the $\{\beta\}$-expansion
is simplified significantly.
Let us show the list of the evident solutions of SE$_6$ taking into account the definition of $Y_4$ in (\ref{def:rY})
\begin{subequations}
 \label{eq:XYZ}
  \begin{eqnarray}
d_4[0,0,1]&=&\left(Y_4(x_{01},y_{01})- Y_4(r_{012})\right)/\beta_2(x_{01},y_{01}),\label{eq:XYZa}\\
d_4[0,1]&=&\left(Y_4(x_{02},y_{02})- Y_4(r_{012})\right)/\beta_1(x_{02},y_{02}), \label{eq:XYZb}\\
d_4[0,1]&=&\left(Y_4(x_{0})- Y_4(r_{012})-\beta_2(x_{0})d_4[0,0,1]\right)/
\beta_1(x_{0}). \label{eq:XYZc}
\end{eqnarray}
 \end{subequations}
 Here Eq.(\ref{eq:XYZc}) together with Eq.(\ref{eq:XYZa}) admit checking of Eq.(\ref{eq:XYZb}).
 The solutions in (\ref{eq:XYZd40},\ref{eq:XYZ}) for $d_4[0], d_4[0,1], d_4[0,0,1]$ look evidently
 easier than ones in (\ref{eq:d_4Sol}) obtained with only single additional d.o.f.
 The solutions for the elements $d_4[1],~d_4[2]$ can also be easily obtained,
 but they look rather cumbersome and we do not show them.
 The solutions presented here exhaust the problem of  fixing  the elements
 of the   $\{\beta\}$-expansion in N$^3$LO.
\section{What can we expect for $\{\beta\}$-expansion in N$^4$LO}
 \label{sec:4}
Let us consider the structure of a 6 loop result in order $a_s^5$, \ie, at $n=5$,
\begin{eqnarray}
 \label{eq:d_5}
\!\!\!\!\!\!\!\!\!d_5(x,\ldots)&=& \beta_0^4\, d_5[4]+ \beta_2\beta_0\, d_1[1,0,1]+ \beta_1^2\,d_5[0,2]+\beta_1 \beta_0^2\,d_5[2,1] + \beta_3\, d_5[0,0,0,1]+\nonumber \\
  & &\beta_0^3\, d_5[3]+
     \beta_1\,\beta_0 d_5[1,1]
     + \beta_2\, d_5[0,0,1] +\nonumber \\
  & & \beta_0^2\, d_5[2]
     + \beta_1\,  d_5[0,1]
     + \beta_0\,d_5[1]+d_5[0].
\end{eqnarray}
The number of new elements in this order,
counting the elements starting with $ \beta_0^4$ up to $\beta_3$ in the first line of Eq.(\ref{eq:d_5}),
coincides with the number of partitions $p(5-1)=5$.
The other terms in (\ref{eq:d_5}) repeat the structure of the result in previous order at $n=4$.
In general, for the term of the order $n$, $d_n$, one should count new terms from $ \beta_0^{(n-1)}$ up to $\beta_{(n-1)-1}$ that gives their number $p(n-1)$,
while the complete number $N(n)$ of all the terms  is the sum
$N(n)=\sum_{l=0}^{(n-1)} p(l)$ that leads to series $\{1, 2, 4, 7, \bm{12}~(\text{here}), 19, 30,45,\ldots \}$ sequentially in each order,
see, \eg, \cite{OEIS}\footnote{I thank N.~Volchanskiy who paid  my attention to this ref.}, and the terms in Eqs.(\ref{eq:d_beta}), for an example.

The coefficient $d_5(x,y)$ is formed by the variety of 6-loop diagrams that get contributions from the intrinsic box- and
pentagon-subgraphs with gluon legs that introduce into $d_5$ a specific $(n_f, n_{\tilde{g}})$-dependence
that does not relate to the charge renormalization.
Indeed, the new color coefficients
$ \Ds
\frac{n_{\tilde{g}}}{2}~d_F^{a b c d e} d_A^{a b c d e}/d_R\,
$ (gluino pentagon inside),
$\, n_f~d_F^{a b c d e}d_F^{a b c d e}/d_R
$ (quark pentagon inside)
enter into $d_5[0](x,y)$ together with the contributions from the box-graphs,
which was already mentioned in Sec.\ref{sec:3}.
The contributions from the latter box-graphs,
$ \Ds
n_{\tilde{g}} d_F^{a b c d} d_A^{a b c d}/d_R\,,$
$~n_f d_F^{a b c d }d_F^{a b c d}/d_R$
enter\footnote{
the definition of the color elements $d^{a_1 a_2\ldots a_n}_{R}$
is presented, \eg, in \cite{Zoller:2016sgq}
}
 now into the element $d_5[1] \to d_5[1](x,y)$.
All these contributions, proportional to $n_f,~n_{\tilde{g}}$,
are well recognized and can be accumulated in the specific term $\delta d_5(x,y)$,
like it was done for $\delta d_4(x,y)$ in Sec.\ref{sec:3}.
The element $\beta_0^4\, d_5[4]$ should also be well recognized;
therefore, one has $11=\bm{12}-1$ unknown elements $d_5[\cdot]$.
By analogy with the previous lower orders procedure one can compile SE$_{11}$
based on Eq.(\ref{eq:d_5}) with the rearranged LHS
$$Y_5(X)=\bar{d}_5(X)-\delta d_5(X) =d_5(X)-\beta_0^4(X)\, d_5[4] - \delta d_5(X)\,,$$
take the equation with the arguments at 11 points ($X_{11}$) on the plane $(x,y)$.
The SE$_{11}$ constructed in this way has unique solution with respect to the $d_5[\cdot]$
elements under the condition the corresponding determinant of the system
$\Delta_{11}(X_{11})\neq 0 $.

Let us take for these 11 components of $X_{11}$ the roots of the equations
\begin{subequations}
\label{eq:11-cond}
\ba
 \!\!&&\!\!\!\!\!\! \beta_0(x_0)=0, ~\beta_1(x_1)=0, ~\beta_2(x_{2m})=0, ~\beta_2(x_{2p})=0, \label{eq:11-cond-a}\\
 \!\!&&\!\!\!\!\!\! \{\beta_0(x_{01},y_{01})=0, \beta_1(x_{01},y_{01})=0\}, \{\beta_0(x_{02m},y_{02m})=0,\beta_2(x_{02m},y_{02m})=0\},\label{eq:11-cond-b} \\
\!\! &&\!\!\!\!\!\!  \{\beta_0(x_{02p},y_{02p})=0,\beta_2(x_{02p},y_{02p})=0\},\label{eq:11-cond-c} \\
\!\! &&\!\!\!\!\!\!
  \{\beta_1(x_{12},y_{12})=0,\beta_2(x_{12},y_{12})=0\}, \{\beta_0(x_{03},y_{03})=0,\beta_3(x_{03},y_{03})=0\}, \\
\!\! &&\!\!\!\!\!\!
  \{\beta_1(x_{13},y_{13})=0,\beta_3(x_{13},y_{13})=0\}, \{\beta_2(x_{23},y_{23})=0,\beta_3(x_{23},y_{23})=0 \}.
  \ea
  \end{subequations}
 This choice of $X_{11}$ simplifies the set of equations, as we made sure in the previous cases of the constructing sets  $X_{3,6}$ in Eqs(\ref{eq:3X}, \ref{eq:6-2-cond}), respectively.
 The determinant corresponding to SE$_{11}$  $\Delta_{11}(X_{11})\neq 0$ but looks too cumbersome
 to show it here.
It is clear that our algebraic scheme works further with the increasing perturbation order.

\section{Conclusion}
In this paper we have considered the important task of obtaining the elements of the detailed structure
of the QCD perturbation expansion -- the $\{\beta \}$-expansion for the renormalization group invariant quantities.
The explicit knowledge of the elements of this expansion (i) gives a possibility to perform various
kinds of optimization of the perturbation series;
(ii) taken together with the Crewther relation it allows one to establish nontrivial
relations between different physical quantities.
 We suggest an algebraic approach to fix the elements of the $\{ \beta \}$-expansion for
these quantities using additional degrees of freedom,
and demonstrate that for the resolution of the detailed structure it is enough
to use a \textit{single additional} degree of freedom to the quark one.

This approach is discussed in detail for N$^2$LO calculations
of the nonsinglet Adler $D$-function and  for the Bjorken polarized sum rules $C^\text{Bjp}$
within QCD with the $n_{\tilde{g}}$ of MSSM gluinos -- the add\-itional degree of freedom.
We derive the explicit formulae for the elements of the $\{ \beta \}$-expansion for these
  quantities, named $d_n[\cdot]$ and $c_n[\cdot]$ respectively,
see Eq.(\ref{eq:d_4})   in the actual case of N$^3$LO within the aforementioned quantum field theory scheme.
This $\{\beta \}$-expansion together with the explicit expressions for the elements $d_4[\cdot]~(c_4[\cdot])$ can be considered as a prediction for any additional
degrees of freedom that can be taken into consideration.
Indeed, these degrees of freedom enter into either the well-known coefficients of the $\beta$-function, $\beta_i$,
or the well-recognized terms of the structure.

  Another kind of predictions is provided by the relation between the elements $d_n[\cdot]$ and $c_n[\cdot]$
  in virtue of the Crewther relation.
  We constructed the fixation procedure also for the case of two additional degrees of freedom.
  Finally, we discussed the  structure and properties of $\{\beta\}$-expansion for higher orders considering the N$^4$LO with the $n_{\tilde{g}}$ of MSSM gluinos  as an example,
  where the expansion elements can be also fixed following to our algebraic procedure.

The next natural step in the development of this investigation would be the calculation of
$D$ or $C^\text{Bjp}$ with the additional degrees of freedom in N$^3$LO.
These results allow one
optimize the corresponding approximation for  the physically important $R_{e^+e^- \to \text{h}}(s)$-ratio or for the
Bjorken polarized sum rules.

\acknowledgments
I would like to thank A. Bednyakov, K. Chetyrkin, A. Grozin, L. Lipatov
for the fruitful discussions, to A. Kataev for the discussion and constructive criticism,
and to the reviewer of this paper for the reasonable, clarifying criticism.
The work  was  supported in part by the BelRFFR--JINR,
grant F16D-004 and by   the Russian Foundation for Basic Research,
Grant No.\ 14-01-00647.

 \appendix
  \section{Explicit formulas for the elements of  $D$ and $C$}%
 \renewcommand{\theequation}{\thesection.\arabic{equation}}
\label{App:A}   \setcounter{equation}{0}
For the Adler function $D^\text{NS}$ the corresponding elements read \cite{Kataev:2010du,Kataev:2014jba}\footnote{we had a missprint in the expression
for $d_3[1]$ in articles \cite{Kataev:2010du,Kataev:2014jba}: in the first parenthesis at $C_A$ should be $\Ds -\frac3{4}$,
see Eq.(\ref{eq:d31}), instead of $\Ds +\frac3{4}$ there. }
\begin{subequations}
\label{eq:d1-4}
 \begin{eqnarray}
d_1&=&3{\rm C_F}; \label{D-11}\\
d_2[1]&=&d_1\left(\frac{11}2-4\zeta_3\right);~~~~~
d_2[0]=d_1\left(\frac{\rm C_A}3-\frac{\rm C_F}2\right); \label{D-21} \\
d_3[2]&=&d_1\left(\frac{302}9-\frac{76}3\zeta_3\right);~d_3[0,1]=d_1\left(\frac{101}{12}-8\zeta_3\right);\label{D-32}\\
d_3[1]&=&
    d_1\left({\rm C_A}\left(-\frac{3}4 + \frac{80}3\zeta_3 -\frac{40}3\zeta_5\right) -
    {\rm C_F}\left(18 + 52\zeta_3 - 80\zeta_5\right) \label{D-31}\right); \label{eq:d31} \\
d_3[0]&=& d_1\Bigg(\left(\frac{523}{36}-
        72 \zeta_3\right){\rm C_A^2}
    +\frac{71}3 {\rm C_A C_F} - \frac{23}{2} {\rm C_F^2}\Bigg)~.  \label{D-30}
\end{eqnarray}
\end{subequations}
\begin{subequations}
 \label{eq:c1-4}
\begin{eqnarray}
c_1&=&-3~{\rm  C_F}; \label{c-11}\\
c_2[1]&=& 2 c_1;
c_2[0]= c_1 \left(\frac{1}{3}{\rm C_A}-\frac{7}{2}{\rm C_F}\right); 
 \label{c-21}
\end{eqnarray}
 \begin{eqnarray}
c_3[2]&=& \frac{115}{18} c_1;~c_3[0,1]=c_1\bigg(\frac{59}{12}-4\zeta_3\bigg);
\label{C-32} \\
c_3[1]&=& -c_1\bigg(\bigg(\frac{166}{9}- \frac{16}3\zeta_3\bigg){\rm C_F} +
\bigg(\frac{215}{36}- 32 \zeta_3+\frac{40}{3}\zeta_5\bigg){\rm C_A}\bigg);
\label{C-31} \\
c_3[0] &=&c_1\bigg(
\bigg(\frac{523}{36} - 72\zeta_3\bigg){\rm C_A^2}+\frac{65}{3}{\rm C_F C_A}+ \frac{\rm C_F^2}{2}\bigg)
\label{C-30}
\end{eqnarray}
\end{subequations}
\be
d_3[0,1] - c_3[0,1]=d_1 \left(\frac{40}3 -12\zeta_3 \right)
\ee
\begin{eqnarray}
 (d_4[0]+c_4[0])(x,y)&=& \tilde{d}_4[0]+\tilde{c}_4[0]= 2d_1 d_3[0]-3d_1^2d_2[0]+d_2[0]^2+d_1^4= \nonumber \\
                        &=& d_1^2\left[ \left(\frac{175}6 -144 \zeta_3 \right) {\rm C_A^2} +
                        44{\rm C_F C_A} - \frac{37}4{\rm C_F^2}\right]
\end{eqnarray}
From the results in \cite{Baikov:2010je} it follows that
\ba
d_4[3]&=&
    d_1\left( \frac{6131}{27}-\frac{406}{3}\zeta_3-60\zeta_5\right),~c_4[3]= c_1\left( \frac{2}{27}\right), \\
\delta d_4(x,y)&=& \left[\frac{y}{2C_\text{A}}\frac{d_A^{a b c d}d_F^{a b c d}}{d_R}+x\frac{d_F^{a b c d}d_F^{a b c d}}{d_R}\right] 3 \cdot \left(- 104-128\zeta_3+ 320 \zeta_5\right),\\
~\delta c_4(x,y)&=&-\delta d_4(x,y). \label{eq:A6}
\ea
\section{The $\beta$-function coefficients}
\label{App:B}   \setcounter{equation}{0}
The required $\beta$-function coefficients with the Minimal Supersymmetric Model (MSSM)\ light
gluinos $n_{\tilde{g}}$ \cite{Clavelli:1996pz},
  and the number $n_f$ of quark flavors, calculated in the \MSbar scheme are
\begin{subequations}
 \label{eq:beta0-3}
\begin{eqnarray}
 \label{eq:beta-b0}
  \beta_0\left(x, y\right)
   &=&\frac{11}{3} C_\text{A} -x-y =\frac{11}{3} C_\text{A} - \frac{4}{3}\left( T_R n_f + \frac{n_{\tilde{g}} C_\text{A}}{2}\right) \,;
\end{eqnarray}
\begin{eqnarray}
 \label{eq:beta-b1}
  \beta_1\left(x, y\right)
     &=& \frac{34}{3}C_\text{A}^2- \left(5 C_\text{A}+3{\rm C_F } \right) x- 8 C_\text{A} y \\
      &=& \frac{34}{3}C_\text{A}^2 - \frac{20}{3}C_\text{A} \left( T_R n_f + \frac{n_{\tilde{g}} C_\text{A}}{2}\right)
      -4\left( T_R n_fC_\text{F}+ \frac{n_{\tilde{g}} C_\text{A}}{2} C_\text{A}\right); \nonumber
\end{eqnarray}
\begin{eqnarray}
 \label{eq:beta-b2}
  \beta_2\left(x, y\right)
      &=& \frac{2857}{54}C_\text{A}^3 -x\left( \frac{1415}{36}C_\text{A}^2 +\frac{205}{12}C_\text{A}C_\text{F} -\frac{3}2C_\text{F}^2  \right) +x^2\left( \frac{11}{4}{\rm C_F } +\frac{79}{24}C_\text{A} \right) - \nonumber \\
       &&\frac{494}{9}C_\text{A}^2 + 2xy \left( \frac{11}{8}C_\text{F} +\frac{14}{3}C_\text{A} \right)C_\text{A}
       +y^2 \frac{145}{24} C_\text{A} ; \\
     &=& \frac{2857}{54}C_\text{A}^3
       - n_f T_R \left( \frac{1415}{27}C_\text{A}^2 +\frac{205}{9}C_\text{A}C_\text{F} -2C_\text{F}^2 \right)
       + (n_f T_R)^2 \left( \frac{44}{9}{\rm C_F } +\frac{158}{27}C_\text{A} \right) - \nonumber \\
       && \frac{988}{27}n_{\tilde{g}} C_\text{A} (C_\text{A}^2) +
       n_{\tilde{g}} C_\text{A} n_f T_R \left( \frac{22}{9}C_\text{A}C_\text{F} +\frac{224}{27}C_\text{A}^2 \right)
       +(n_{\tilde{g}} C_\text{A})^2 \frac{145}{54} C_\text{A} \, , \nonumber
\end{eqnarray}
 \end{subequations}
 where we have introduced appropriate rescaled variables $\Ds x= \frac{4}{3} T_R n_f ~~\text{and}~y=\frac{4}3\frac{C_\text{A}}2 n_{\tilde{g}}$ after the first equality to simplify the expressions.
 The N$^3$LO coefficient $\beta_3\left(n_f, n_{\tilde{g}}\right)$ has been obtained recently in  \cite{Zoller:2016sgq,Bednyakov:2016uia},
\begin{eqnarray}
\beta_{3}\left(n_f, n_{\tilde{g}}\right)&=&
          - \left( \frac{150653}{486} - \frac{44}{9} \zeta_{3}\right) C_\text{A}^4          + \left(\frac{80}{9} - \frac{704}{3} \zeta_{3} \right)\dAANA \nonumber\\ & &
          +\Ngl\left[ \left( \frac{68507}{243} - \frac{52}{9} \zeta_{3} \right)C_\text{A}^4 - \left(\frac{256}{9} - \frac{832}{3} \zeta_{3}\right)\dAANA \right]  \nonumber \\
&& -\Ngl^2\left[\left( \frac{26555}{486} - \frac{8}{9} \zeta_{3} \right)C_\text{A}^4 + \left(\frac{176}{9} - \frac{128}{3} \zeta_{3}\right)\dAANA \right]
          - \frac{23}{27} C_\text{A}(\Ngl C_\text{A})^3 \nonumber
\end{eqnarray}
\begin{eqnarray}
\phantom{\beta_{3}\left(n_f, n_{\tilde{g}}\right)}&&
          + \Nf \tr \left[
          - 46  C_\text{F}^3
          + \left(\frac{4204}{27} - \frac{352}{9} \zeta_{3}\right) C_\text{A} C_\text{F}^2
          - \left(\frac{7073}{243} - \frac{656}{9} \zeta_{3}\right)C_\text{A}^2 C_\text{F} \right. \nonumber\\ & &\left.
          + \left(\frac{39143}{81} - \frac{136}{3} \zeta_{3}\right) C_\text{A}^3
          \right]
          -  \Nf \left(\frac{512}{9} - \frac{1664}{3}\zeta_{3}\right) \dRANA  \nonumber\\ & &
          +(\Nf \tr)^2 \left[
          - \left(\frac{1352}{27} - \frac{704}{9} \zeta_{3}  \right)C_\text{F}^2
          - \left(\frac{17152}{243} + \frac{448}{9} \zeta_{3} \right)C_\text{A}  C_\text{F} \right. \nonumber\\
          & &\left.
          - \left(\frac{7930}{81}  + \frac{224}{9} \zeta_{3} \right)C_\text{A}^2
          \right]
          + \Nf^2 \left(\frac{704}{9}  - \frac{512}{3} \zeta_{3}\right) \dRRNA  \nonumber\\
         & &
          -(\Nf \tr)^3 \left[
           \frac{1232}{243} C_\text{F}
          + \frac{424}{243}  C_\text{A}
          \right] \nonumber
          \end{eqnarray}
\begin{eqnarray}
        & &
          +\Ngl C_\text{A} (\Nf \tr) \left[
           \left(\frac{152}{27} + \frac{64}{9} \zeta_{3}\right) C_\text{F}^2
          - \left(\frac{23480}{243}  - \frac{352}{9} \zeta_{3}\right) C_\text{A} C_\text{F} \right. \nonumber\\
          & &\left.
          - \left(\frac{30998}{243} + \frac{128}{3} \zeta_{3}\right) C_\text{A}^2
          \right]
           +\Nf \Ngl  \left(\frac{704}{9} - \frac{512}{3}  \zeta_{3}\right) \dRANA \nonumber\\ & &
          - (\Ngl C_\text{A})^2 \Nf\tr \left[
            \frac{308}{243} C_\text{F}
          + \frac{934}{243} C_\text{A} \right]
          -\Ngl C_\text{A} (\Nf \tr)^2 \left[
           \frac{1232}{243} C_\text{F}
          + \frac{1252}{243} C_\text{A} \right]\,, \label{4lbetaasgl}
\end{eqnarray}
where $\Ds T_\text{R} = \frac{1}{2},\, C_\text{F} = \frac{N^2_c-1}{2N_c},\,  C_A =N_c,\, N_\text{A}=2 C_\text{F} C_\text{A} =N_c^2-1$.

\end{document}